\documentstyle[aps,twocolumn]{revtex}

\begin{document}
\title{Asymptotic normalization coefficient of $^{8}$B from breakup reactions and
the S$_{17}$ astrophysical factor }
\author{L. Trache$^{1}$, F. Carstoiu$^{2,3}$, C. A. Gagliardi$^{1}$, R. E. Tribble$%
^{1}$}
\address{$^{1}$Cyclotron Institute, Texas A\&M University, College Station, TX-77843,%
\\
USA}
\address{$^{2}$Institute of Physics and Nuclear Engineering H. Hulubei, Bucharest,\\
Romania\\
$^{3}$Laboratoire de Physique Corpusculaire, F-14050 Caen Cedex, France}
\date{\today}
\maketitle

\begin{abstract}
We show that asymptotic normalization coefficients can be extracted from one
nucleon breakup reactions of loosely bound nuclei at 30-300 MeV/u. In
particular, the breakup of $^{8}$B is described in terms of an extended
Glauber model. The $^{8}$B ANC extracted from breakup data at several
energies and on different targets, $C_{tot}^{2}=0.450\pm 0.039$ fm$^{-1}$,
leads to the astrophysical factor $S_{17}(0)=$ $17.4\pm 1.5$\ \ eV$\cdot $b
for the key reaction for solar neutrino production $^{7}$Be(p,$\gamma $)$%
^{8} $B. The procedure described provides an indirect method to determine
reaction rates of astrophysical interest with beams of loosely bound
radioactive nuclei.
\end{abstract}


$^{8}$B, produced in the $^{7}$Be($p$,$\gamma $)$^{8}$B reaction, is the
major source of the high-energy neutrinos observed by solar neutrino
detectors\cite{bahcall}. Results of recent direct and indirect
determinations of the astrophysical factor $S_{17}(0)$ (see \cite
{ham,junghans,iwasa,davids,azhari3} and references therein) do not fully
agree, but span the full adopted range, 19$_{-2}^{+4}$ eV\thinspace b \cite
{adelb}. The asymptotic normalization coefficient (ANC) approach \cite{xu}
is based on the fact that the $^{7}$Be($p$,$\gamma $)$^{8}$B reaction is
highly peripheral. Thus, what one needs most is a precise quantitative
description of the tail or periphery of the $^{7}$Be + $p$ $\leftrightarrow $
$^{8}$B overlap function, which is given by the corresponding ANC, rather
than of the full many-body wave function inside the core. The advantage of
this approach is that ANCs may be determined from other peripheral nuclear
reactions that have cross sections orders of magnitude larger than the
radiative capture reaction of interest. This technique has been applied to
determine $S_{17}$ from measurements of two ($^{7}$Be,$^{8}$B) proton
transfer reactions \cite{azhari3}, after being tested to give results to
better than 9\% \cite{gagl99} for $^{16}$O(p,$\gamma $)$^{17}$F.

In the present paper we show that one can extract asymptotic normalization
coefficients for the $^{8}$B ground state wave function from data on
one-proton removal (or breakup) reactions at energies of 30-300 MeV/u \cite
{nego,blank,pecina,kelley}. Improvements in the nuclear reaction model used
make a quantitative description possible for the energy range and the ANC
extracted reliable. Then, this ANC can be used to obtain an independent
determination of the astrophysical factor $S_{17}(0)$.

Recent advances in experimental techniques permit the selection of the final
state of the remaining core in breakup reactions at large energies.
Typically an exotic nucleus $B=(Ap)$, where $B$ is a bound state of the core 
$A$ and the proton $p$, is produced by fragmentation from a primary beam,
separated and used to bombard a secondary target. After the breakup occurs,
the breakup cross section and the parallel momentum distribution of the core 
$A$ are measured. By detecting the final state of the core, e.g. using
coincident gamma rays, one can determine important spectroscopic information
about the ground state of the exotic projectile \cite{motob,navin}. The
measured momentum distribution of the core can be related to the momentum
distribution of the bound nucleon.

In all reactions where the core survives (either transfer or one-nucleon
breakup) the matrix elements include the overlap integral $I_{Ap}^{B}(\vec{r}%
)$ for the nuclei $A$, $p$, and $B$, obtained after the integration over the
internal coordinates of fully antisymmetric wave functions, with $\vec{r}$
the vector connecting the center of mass of nucleus $A$ with $p$ \cite
{satch,xu}. The overlap integrals are not normalized to unity, but to the
spectroscopic factors $S_{nlj}$. At asymptotic distances where nuclear
forces are vanishingly small, $r>R_{N}$, the overlap integrals behave as 
\begin{equation}
I_{Aplj}^{B}(r)\rightarrow S_{nlj}^{1/2}\varphi _{nlj}(r)\rightarrow
C_{Aplj}^{B}\frac{W_{-\eta ,l+1/2}(2kr)}{r}.  \label{asymp}
\end{equation}
Here $C_{Aplj}^{B}$ is the asymptotic normalization coefficient defining the
amplitude of the tail of the overlap integral, $W$ is the Whittaker
function, $k$ is the wave number, and $\eta $ is the Sommerfeld parameter
for the bound state ($Ap$). The asymptotic normalization coefficients $%
C_{Aplj}^{B}$ can be extracted from any peripheral observables that are
measured experimentally.

For the reaction model calculations we assume that the ground state of the
projectile $(J^{\pi })$ can be approximated by a superposition of
configurations of the form $[\Psi _{c}^{\pi _{c}}\otimes nlj]^{J^{\pi }}$,
where $\Psi _{c}^{\pi _{c}}$ denote the core states and $nlj$ are the
quantum numbers for the single particle wave function $\varphi _{nlj}(r)$\
in a spherical mean field potential. These single particle states are
normalized to unity and have the asymptotic behavior given by Eq. \ref{asymp}%
, with the single particle asymptotic normalization coefficients $b_{nlj}$.
When more than one configuration contributes to a selected core state, the
total cross section for one-nucleon breakup is written as an incoherent
superposition of single particle cross sections: 
\begin{equation}
\sigma _{-1p}=\sum S(c,nlj)\sigma _{sp}(nlj).  \label{cs-tot}
\end{equation}
A similar relation holds for the momentum distribution. Typically the
nucleon is not measured, therefore the calculated cross sections $\sigma
_{sp}(nlj)$ contain a stripping term (the loosely bound nucleon is absorbed
by the target and the core is scattered and detected), a diffraction
dissociation term (the nucleon is scattered away by the target, the core is
scattered by the target and is detected) and a Coulomb dissociation term: 
\begin{equation}
\sigma _{sp}=\int_{0}^{\infty }2\pi bdb(P_{str}(b)+P_{diff}(b))+\sigma
_{Coul}  \label{sig-sp}
\end{equation}
In previous analyses of breakup reactions, a structure for the projectile
(the spectroscopic factors $S(c,nlj)$) was assumed and agreement between the
calculated and experimental values was considered a validation both of the
assumed nuclear structure and of the reaction model calculations used \cite
{blank,khal,esben}. In the present case we use an extended Glauber model, in
the eikonal approximation with non-eikonal correction terms up to the second
order \cite{nego,carst-glaub}. Realistic nucleon-target and core-target
S-matrix elements are used in the evaluation of the impact parameter
dependent probabilities. We find that the largest contributions to the cross
sections $\sigma _{sp}(nlj)$ come from large impact parameters and therefore
the phenomena are peripheral. It then follows that we can express the
results in terms of the asymptotic normalization coefficients and reverse
the process: we can use the experimental results to extract the ANCs.

Calculations were done for several $^{8}$B breakup reactions for which data
exist in the literature. The model is similar to that developed by Bertsch
et al. \cite{bertsch,henck}. It has been tested before on 23 different
reactions in\ the {\it p-sd} shell \cite{nego,sauvan}. The loosely bound
proton and the core moving on an eikonal trajectory interact independently
with the target nucleus, an assumption valid at these energies. For the
proton-target interaction we used that of Jeukenne, Lejeune and Mahaux (JLM) 
\cite{jlm}, in the updated version of Bauge et al. \cite{bauge}. For the
target-core nucleus-nucleus interaction we use the double folding procedure
described in \cite{tra-el}: the same JLM interaction is folded with
Hartree-Fock nuclear matter distributions of the core and of the target.
Subsequently the double folding potentials were renormalized to reproduce a
variety of elastic scattering data for light nuclei. We found there that the
real part needed a substantial renormalization at about 10 MeV/u (N$_{V}$%
=0.366), but the imaginary part did not (N$_{W}$=1.00). We have checked the
procedure on a much wider set of data from literature at higher energies,
and found a similar conclusion for the imaginary part, while the
renormalization of the real part approaches unity around 50 MeV/u.
Therefore, in the present calculations we adopted the procedure of \cite
{tra-el}, with the JLM(1) interaction, and N$_{W}$=1.00. The S-matrix
calculations that enter the first two terms of Eq. \ref{sig-sp} are expected
to depend primarily on the imaginary part of the interaction. Indeed,
calculations with large variations of the renormalization of the real
potential (N$_{V}$=0.366 and N$_{V}$=0.80) give the same cross sections. The
integrated Coulomb term is treated in a perturbative method that retains the
dipole and quadrupole terms, equivalent with that of Ref. \cite{baur}, but
using radial matrix elements calculated with realistic Woods-Saxon radial
wave functions.

The impact parameter dependence of the first two terms in Eq. \ref{sig-sp}\
is plotted in Fig. \ref{fig-P(b)} for the case of the breakup of $^{8}$B on
a Si target at 38 MeV/u \cite{nego}. Clearly, both terms are dominated by
the periphery but have contributions from small impact parameters. The
Coulomb term is even more peripheral, due to its long range. To investigate
the influence of the nuclear interior on the extracted ANC, different single
particle wave functions were used for the outer proton to calculate the
breakup cross section in the same reaction model. We chose a range of radii
and diffusenesses ($R=2.20-2.60$ fm and $a=0.50-0.70$ fm) for the
Woods-Saxon potentials and repeated the calculations. A correct spin-orbit
term was included, and in all cases we adjusted the depth of the potential
to reproduce the proton separation energy $S_{p}=137$ keV. The radial
behavior of $1p_{1/2}$ and $1p_{3/2}$ orbitals is identical at large
distances and, for a given ($R,a$), differs at small radii by much less than
the variation associated with the choice of Woods-Saxon potential. Thus, for
simplicity, only the $1p_{3/2}$ component was included, and we rewrite Eq. 
\ref{cs-tot} as: 
\begin{equation}
\sigma _{-1p}=(S_{p_{3/2}}+S_{p_{1/2}})\sigma
_{sp}=(C_{p_{3/2}}^{2}+C_{p_{1/2}}^{2})\sigma _{sp}/b_{p}^{2},  \label{SC2}
\end{equation}
where $b_{p}$ is the ANC of the normalized $1p_{j}$ radial single particle
wave function.\ The experimental value for the breakup cross section is$\
222\pm 15$ mb. The calculated cross section varies from 226 mb to 326 mb
with the choice of the single particle wave function used (i.e. $(R,a)$ the
radius and diffuseness parameters of the Woods-Saxon potential used), which
is equivalent to $44\%$ variation if a spectroscopic factor $%
S_{tot}=S_{p_{3/2}}+S_{p_{1/2}}$ is extracted using the first part of Eq. 
\ref{SC2}. However if the square of the ANC $%
C_{tot}^{2}=C_{p_{3/2}}^{2}+C_{p_{1/2}}^{2}$ is extracted instead, the
result is very stable, as shown in Fig. \ref{fig-SC2} (where for convenience
the results are plotted against the single particle ANC $b_{p}$ calculated
for each geometry assumed). The variation in the ANC over the full range
considered here is 
\mbox{$<$}%
$\pm 3\%$. A flat curve for $C^{2}$ in Fig. \ref{fig-SC2} would be a
signature of a purely peripheral reaction; the small slope reflects the
participation of the interior of the nucleus. Similar calculations were done
for the same target and two other energies (35 and 28 MeV/u), and for other
targets ($^{12}$C, Sn and $^{208}$Pb) at other energies: 40, 142 and 285
MeV/u \cite{pecina,blank}. Pictures similar to Fig. \ref{fig-SC2} are
obtained in each and every case. These experiments have not determined the
yield to the $^{7}$Be first excited state. A Coulomb breakup experiment
carried out at 50 MeV/u on a Pb target \cite{motob} found it to be small,
about 5\% of the total. From this we estimated the ANC for the core
excitation part in the wave function, then calculated its contribution to
the one-proton removal cross section on each target and subtracted it (e.g. $%
7.5\%$ for the Si target). The corrected ANCs extracted are presented in
Table \ref{tab-lst}. With the exception of two data points on $^{12}$C,
there is good agreement among the ANCs which come from data over a wide
range of incident energies and both low and high $Z$ targets. In order to
extract an average ANC, we have done an unweighted average of the individual
measurements. Using all of the data points, this results in $%
C_{tot}^{2}=0.450(30)$ fm$^{-1}$ where the uncertainty is the standard error
of the mean. If the two $^{12}$C data points (4 and 5 in the table) at 40
and 142 MeV/u are removed from our average (note that they fail a simple
test of the expected energy dependence), we find $C_{tot}^{2}=0.456(14)$ fm$%
^{-1}$. Several correlated uncertainties must be added: 4\% for the
renormalization coefficients used for the optical model parameters, 3\% for
the variation in $C^{2}$ as a function of $b_{p}$ and 2\% for the
uncertainty in the excited state contribution. Including these we find $%
C_{tot}^{2}=0.450(39)$ fm$^{-1}$ for the full data set and $%
C_{tot}^{2}=0.456(28)$ fm$^{-1}$ when the two $^{12}$C data points are
removed. The values are similar and, without additional information, we
shall adopt the first.

Using a wave function with the asymptotic normalization as extracted above,
we have calculated the distribution of the parallel momentum of the core
measured in the breakup of $^{8}$B at 41 MeV/u on a $^{9}$Be target \cite
{kelley}. The result is compared in Fig. \ref{fig-mdb} with the experimental
data. A very good description is obtained. Notably, unlike the black disk
model \cite{hansen} calculation which describes the width of the parallel
momentum distribution \cite{carb8}, the extended Glauber model calculation
also matches the large momentum tails due to the nuclear interior. This
gives further credibility to our calculations and the entire approach.

The value extracted here from the breakup of $^{8}$B at 30-300 MeV/u agrees
very well with the one extracted from transfer reactions at 12 MeV/u \cite
{azhari3}. We can use the ANC extracted to evaluate the astrophysical S
factor for the reaction $^{7}$Be(p,$\gamma $)$^{8}$B at very low energies,
following the procedure of Ref. \cite{xu}. Using the value from breakup we
find $S_{17}(0)=17.4\pm 1.5$ eV$\cdot $b. We can also use the ANC extracted
here to determine the rms radius for the $^{8}$B proton halo, following the
procedure of \cite{carb8}. We find a $r_{h}=4.20\pm 0.21$ fm.

In conclusion, reliable spectroscopic information can be extracted from
one-nucleon breakup reactions of loosely bound nuclei at energies around and
above the Fermi energy. However, we have shown that, despite a more
transparent meaning of the spectroscopic factors, the values obtained are
not unambiguous, and a better quantitative description is achieved in terms
of the asymptotic normalization coefficients. In turn, these can be used to
calculate any observables that are dominated by the periphery of the
nucleus, notably rms radii for halos and astrophysical S-factors.
Calculations using an extended Glauber model for the breakup data of $^{8}$B
on a wide range of targets and energies lead to an unambiguous value for the
ANC and an astrophysical factor $S_{17}(0)$ in very good agreement with the
values from recent determinations from direct measurements \cite{ham} and
with those using indirect methods \cite{davids,azhari3}. New measurements
for the elastic scattering of $^{8}$B, a more accurate determination of the
breakup cross sections (eventually separating the stripping and diffraction
dissociation components) and a precise determination of the core excitation
contribution, can increase the reliability of the ANC extracted. The
validity of the procedure is wider than for the $^{8}$B case discussed
above. In addition to peripherality, ensured more or less for the halo
nuclei, the requirements are good absolute values for the breakup cross
sections, with the identification of the final state of the core, and
reliable cross section calculations. The method can be used to extract
valuable information for nuclear astrophysics. Very difficult or even
impossible direct measurements that would involve bombarding short lived
targets with very low energy protons can be replaced or supplemented by
indirect methods seeking the relevant ANCs, rather than complete knowledge
of the ground state wave function of these exotic nuclei. In addition, the
indirect ANC method is subject to different systematic errors than the
direct measurements, and therefore redundance of the results is very much
welcome, particularly for critical astrophysical S-factors, such as that for
the $^{7}$Be(p,$\gamma $)$^{8}$B reaction.

One of us (FC) acknowledges the support of IN2P3 for a stay at Laboratoire
de Physique Corpusculaire in Caen, during which part of this work was
completed. The work was supported by the U. S. Department of Energy under
Grant No. DE-FG03-93ER40773, by the Romanian Ministry for Research and
Education under contract no 555/2000, and by the Robert A. Welch Foundation.

\begin{figure}[tbp]
\caption{ The stripping and diffraction dissociation parts of the breakup
probability as a function of the impact parameter. }
\label{fig-P(b)}
\end{figure}

\begin{figure}[tbp]
\caption{ Comparison of the spectroscopic factors S$_{tot}$(dots) and of the
ANC $C_{tot}^{2}$ (triangles) extracted from the $^{8}$B breakup data on Si
at 38 MeV/u \protect\cite{nego}, for different parameters of the single
particle Woods-Saxon potentials. }
\label{fig-SC2}
\end{figure}

\begin{figure}[tbp]
\caption{ Calculated parallel momentum distributions for core-like fragments
in a breakup reaction of $^{8}$B on a $^{9}$Be target at 41 MeV/u, are
compared with experimental data \protect\cite{kelley}. The curve labeled
''intr'' is the result of a calculation with the Serber model in the
transparent limit, that labeled ''disk'' with the black disk model. }
\label{fig-mdb}
\end{figure}

\begin{center}
\begin{table}[tbp]
\caption{ Summary of the ANC extracted from different $^{8}$B breakup
reactions. }
\label{tab-lst}
\end{table}
\begin{tabular}{|l|l|l|l|l|}
\hline
Target & E/A & exp. c.s. & Ref. & $C_{tot}^{2}$ \\ 
& [MeV/u] & [mb] &  & [fm$^{-1}$] \\ \hline
$^{28}$Si & 28 & 244(15) & \cite{nego} & 0.435 \\ 
& 35 & 225(15) & \cite{nego} & 0.420 \\ 
& 38 & 222(15) & \cite{nego} & 0.423 \\ 
$^{12}$C & 40 & 80(15) & \cite{pecina} & 0.250 \\ 
& 142 & 109(1) & \cite{blank} & 0.597 \\ 
& 285 & 89(2) & \cite{blank} & 0.482 \\ 
Sn & 142 & 502(6) & \cite{blank} & 0.547 \\ 
& 285 & 332(6) & \cite{blank} & 0.464 \\ 
$^{208}$Pb & 142 & 744(9) & \cite{blank} & 0.421 \\ 
& 285 & 542(9) & \cite{blank} & 0.460 \\ 
aver &  &  &  & 0.450(30) \\ \hline
\end{tabular}
\end{center}

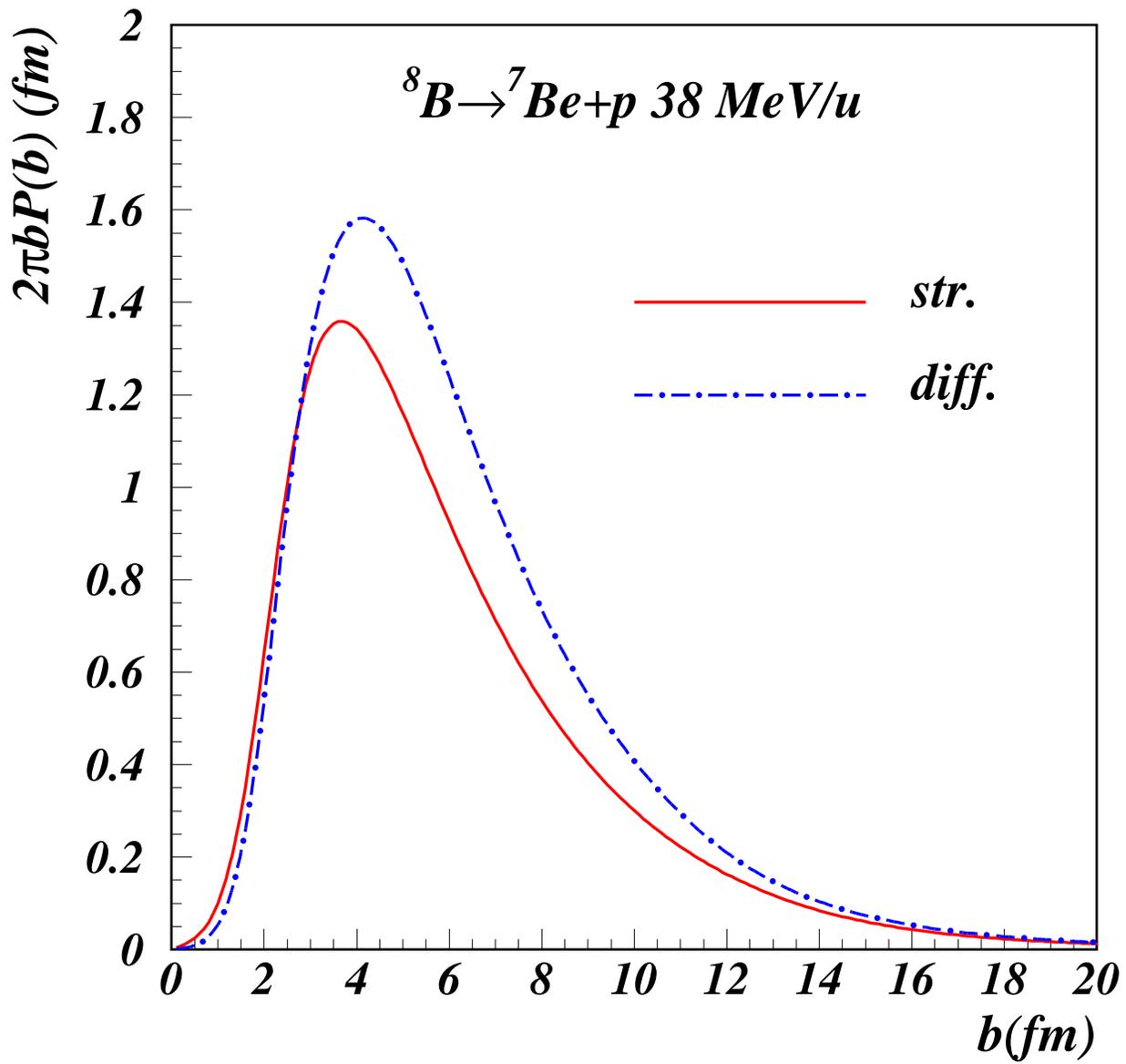
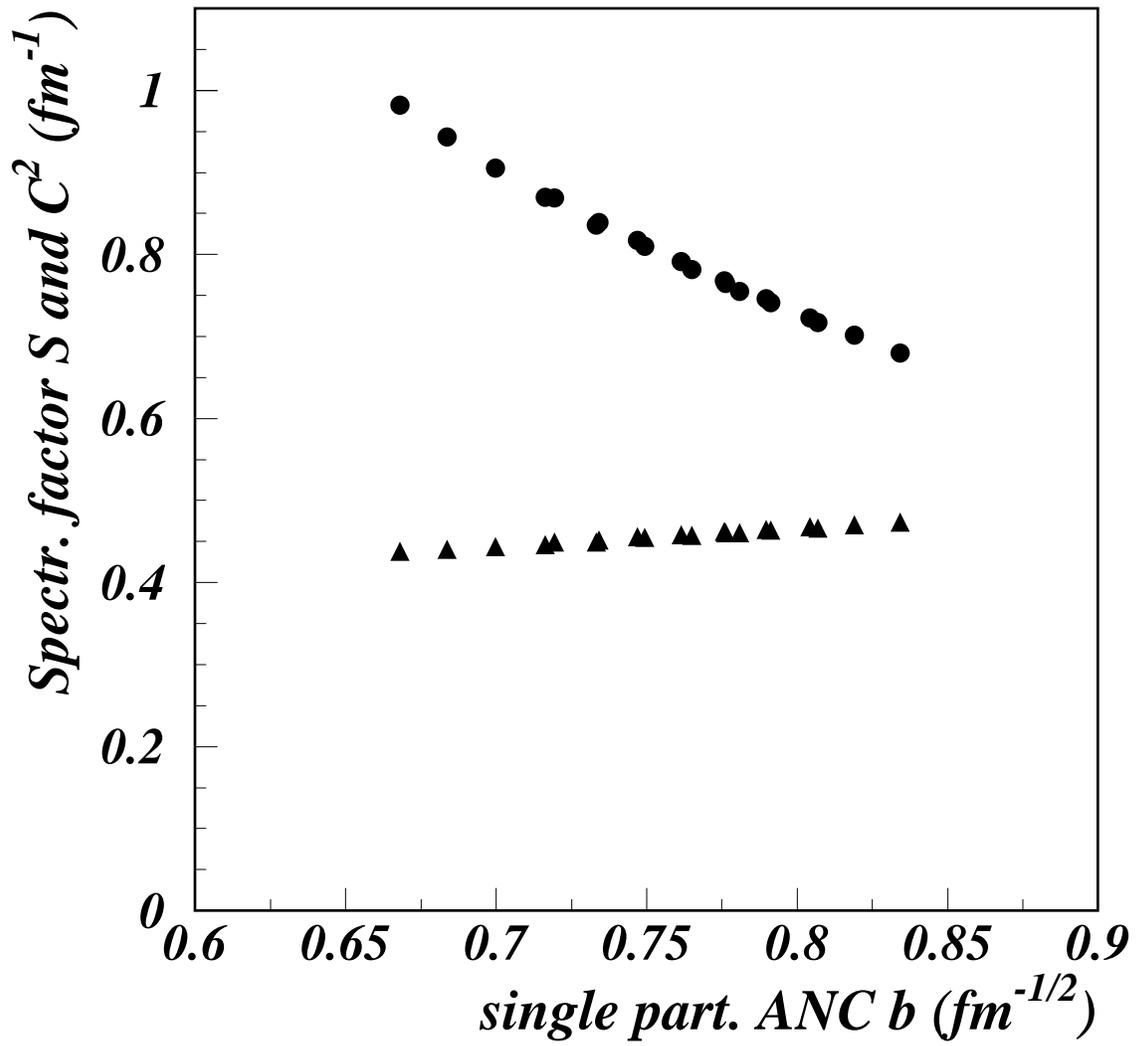
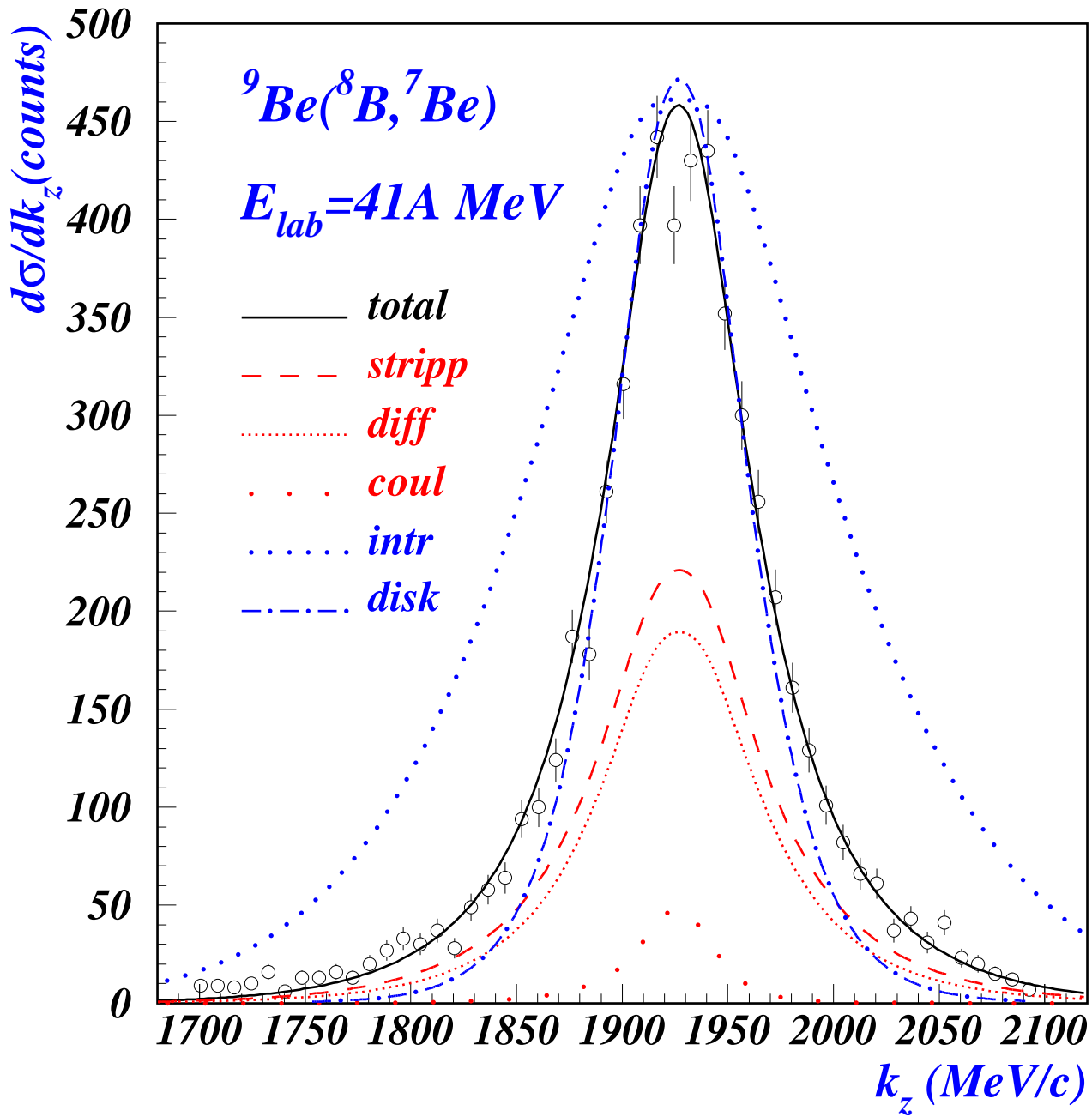
\end{document}